# Does Aesthetics of Web Page Interface Matters to Mandarin Learning?

***Jasni Mohamad Zain†, Mengkar Tey††, and Yingsoon Goh†††***

***† and ††Faculty of Computer Systems and Software Engineering, University Pahang Malaysia, Karung Berkunci 12, 25000 Kuantan, Pahang, Malaysia.***
***†††Academy of Language Studies, UiTM Terengganu, 23000 Dungun, Terengganu, Malaysia.***

**Summary**
Aesthetics of web page refers to how attractive a web page is in which it catches the attention of the user to read through the information. In addition, the visual appearance is important in getting attentions of the users. Moreover, it was found that those screens, which were perceived as aesthetically pleasing, were having a better usability. Usability might be a strong basic in relating to the applicability for learning, and in this study pertaining to Mandarin learning. It was also found that aesthetically pleasing layouts of web page would motivate students in Mandarin learning The Mandarin Learning web pages were manipulated according to the desired aesthetic measurements. GUI aesthetic measuring method was used for this purpose. The Aesthetics-Measurement Application (AMA) accomplished with six aesthetic measures was developed and used. On top of it, questionnaires were distributed to the users to gather information on the students' perceptions on the aesthetic aspects and learning aspects. Respondents for this study were students taking Mandarin course level I at UiTM Terengganu. A significant correlation of the aesthetic aspect was found with its relevance to Mandarin learning. In summary, aesthetics should not be ignored or overlooked in designing effective learning interfaces for educational purposes.

***Key words:***
*Aesthetics, Usability, Aesthetic Measures, Mandarin Learning Web Page, Mandarin Learning*

## 1. Introduction

Aesthetics of web pages refers to how attractive a web page is in which it catches the attention of the user to read through the information. Visual appearance is important in getting attentions of the users to browse through the entire web pages. It was found that those screens, which were perceived as aesthetically pleasing, were having a better usability [11]. In addition, aesthetic measuring methods of web page interface are important as this may help in gaining students' attention and in erecting their interest in using the interface However, aesthetics of the web page interfaces can be very subjective. Web page interface was a kind of Graphical User Interface (GUI), and the term "aesthetics of GUIs" is so subjective. This is because different people have different views and opinions. It is very difficult to comment or conclude which interface is the most beautiful or ugly.

Therefore, it is normally very difficult to judge whether an interface is exquisite or not. Consequently, the major concern of this research is to provide an objective tool for unbiased aesthetics measurement whereby the users' perceptions are supposed to be congruent with the aesthetic values of the Aesthetics-Measurement Application (AMA) used. This is to ease the measurement of aesthetics of Mandarin learning web pages particularly and web pages of any languages generally in the future. Some research showed that an important aspect of screen design is aesthetic evaluation of screen layouts. A very essential component of GUI design involves the actual layout of elements on the screen. In this research, GUI aesthetic measuring methods are important to prospective viewers as this may help gain students' attention and build their confidence in using the interface.

In order to achieve the above-mentioned purpose, Mandarin learning web pages used for this research comprise of main pages, learning pages and exercise pages that were manipulated according to the desired aesthetic values. The six aesthetics related elements used are balance, equilibrium, symmetry, sequence, rhythm, as well as order and complexity. With this intention, the study was held to gather students' views on the aesthetic aspects. It was found that the users' perceptions were congruent with the aesthetics values gathered by using our self-developed Aesthetics-Measurement Application (AMA). Hence, our AMA perhaps can be introduced as an effortless tool for web page aesthetics measurement.

Online learning via web page in language teaching and learning gains its popularity in the teaching and learning of Mandarin. It is important that web page for learning language, especially Mandarin, can be designed aesthetically. This is because aesthetically pleasing layouts of web page will motivate students or users to learn Mandarin. Therefore, learning pages with high aesthetic value as learning supplementary materials can inspire students towards learning and make learning process easy. Thus, it is the purpose of this article to explore the relationship of aesthetics with Mandarin learning specifically and the relationship of aesthetics with learning in general. We believe aesthetics may bring impact on learning.

## 2. Literature Reviews

There were limited literature reviews concerning the aspect of aesthetics on learning. Nevertheless, we believe that a well-designed screen could increase human processing speed, reduce human errors, and speed up computer-processing time and thus it will increase human productivity and usability. It was found that the result of the findings on aesthetic aspect of Mandarin web pages for users' perceptions were congruent with the aesthetics values gathered by using Aesthetics Measurement Application [15].

We need to find ways in measuring aesthetics, as aesthetics might be a subjective concept. Efforts have been carried out for aesthetics measurements. Approaches and ways of measuring aesthetics were conducted. Graphics design experts derived a number of principles for what comprised an aesthetic design [1][13]. These principles included balance, regularity, symmetry, predictability, economy, sequentially, unity, proportion, simplicity, and grouping. In addition to these, the calculations and technique of aesthetics measurement for web page interface were derived from the past research that focused on fourteen aesthetic measurements elements, balance, equilibrium, symmetry, sequence, cohesion, unity, proportion, simplicity, density, regularity, economy, homogeneity, rhythm, as well as order and complexity [2]. Some of the elements were alike in the comparison with Taylor's (1960) and Dondis (1973)'s measurements.

It was agreed that, symmetry was one of the most fundamental principles in design and it would affect the layout and feeling of a design as a symmetrical page might give a feeling of permanence and stability but asymmetrical balance might impel interest and should not be neglected [4]. Prior studies on the aesthetic aspects of interfaces focused prematurely on visual design elements as the objects [6]. The visual design elements posed in a graphical user interface would be regarded as objects that have effect on the aesthetic quality of a GUI.

It was found that aesthetic responses were closely related to usability [7][16]. There was a positive correlation between the aesthetic aspects and perceived usability of Virtual Learning Environments (VLE) interfaces [10]. Those screens, which were perceived as aesthetically pleasing, were having a better usability [11]. On one other hand, the texts and graphics used in the e-portfolio in aesthetic quality aspect is an aspect that ought not to neglect because the attractiveness of the e-portfolio would be an affirmative factor that drawn students in the use of e-portfolio [2].

In addition, good aesthetic layouts definitely affect a student's motivation to learn [15], as related to ARCS model [5]. The ARCS model refers to Attention, Relevance, Confidence, and Satisfaction, and its explanations are shown in Table 1. It was found that aesthetically pleasing layouts of web page would motivate students in Mandarin learning [14].

Table 1: ARCS Model

| | Model | Explanation |
|---|---|---|
| **A** | **Attention** | Good layouts will attract the attention of the student. |
| **R** | **Relevance** | Good layouts will be relevant to the student. |
| **C** | **Confidence** | Good layouts will boost the student's confidence. |
| **S** | **Satisfaction** | The student will feel satisfied if the design is good and appealing. |

On top of these, few researches have been conducted on applying aesthetics to interface designs [8][9][14]. In our paper, the aesthetic values of interface designs were manipulated in order to investigate their effect on learning.

In order to achieve the above mentioned purpose, ASSURE model as shown in the Table 2, was introduced in this paper helpful for designing learning lessons using different kinds of media [3]. This model assumed that instructions were not delivered using lecture or textbook only, it allowed for the possibility of incorporating out-of-class resources and technology into the class materials and would be especially helpful for instructors using online learning materials. This included all kinds of GUI where students were required to interact with the provided learning materials. ASSURE model was a procedural guide for planning and conducting instruction that incorporates media and technology where learning occurred [12].

Table 2: ASSURE Model

| Model | Explanation |
|---|---|
| **A** | Analyze Students |
| **S** | State Objectives |
| **S** | Select Media and Materials |
| **U** | Utilize Media and Materials |
| **R** | Require Learner Participation |
| **E** | Evaluate and Revise |

## 3. Approach and Methodology

This research was carried out in two major parts. The first part concentrated on GUI aesthetics measurement by using our self-developed Aesthetics-Measurement Application (AMA). The second part was User's Perceptions of Aesthetics on web page Interfaces and Aesthetics relevance with Mandarin Learning. This research focuses only on the aesthetics of the positions of objects, images element and texts element as objects in a multi screen interface along with its' relevance to Mandarin learning.

### 3.1 Self-developed Aesthetics-Measurement Application (AMA)

Figure 1 showed a self-developed Aesthetics-Measurement Application (AMA) and the aesthetic value counted on the Main page interface of the Mandarin Learning web pages. AMA was developed by using Matlab software based on six aesthetic measures derived from the model of Ngo, Teo, and Byrne (2003). Table 3 showed the explanations of terms used and the formulae of six aesthetic measures used for aesthetics measurement. The six elements involved were balance, equilibrium, symmetry, sequence, rhythm, as well as order and complexity. The definitions of terms as well as the mathematical formulae were included.

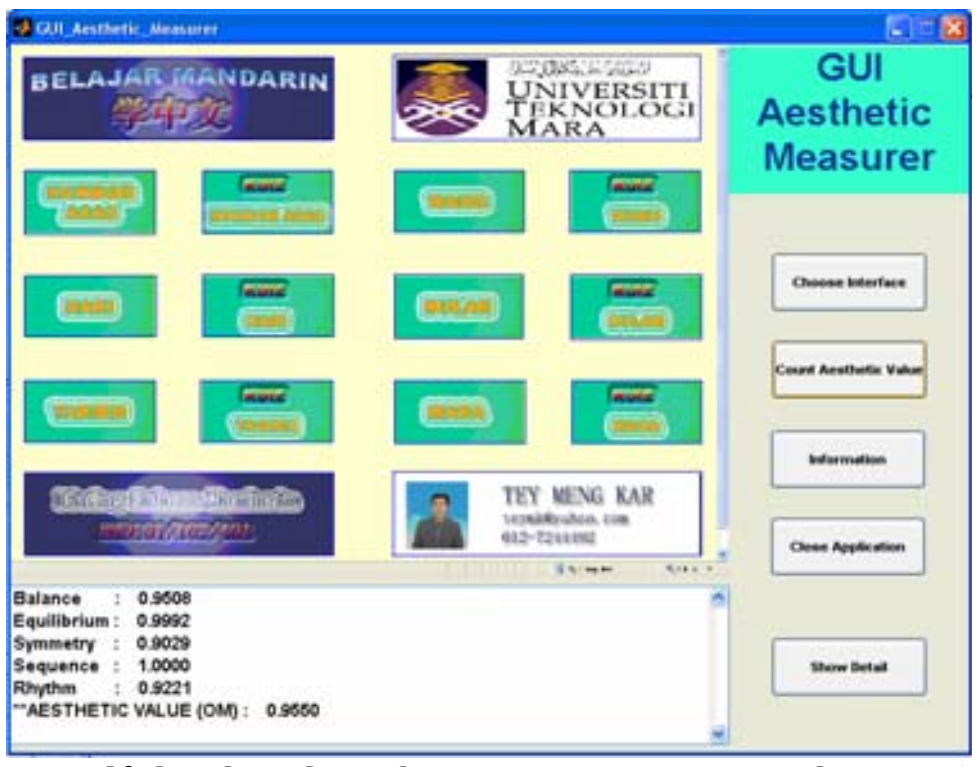

Figure 1: Self-developed Aesthetics-Measurement Application (AMA)

Table 3: Six Aesthetic Measures Used for Aesthetics Measurement
Source: Ngo, Teo, and Byrne (2003)

1. Balance

$$Balance = 1 - \frac{|BalanceVer| + |BalanceHor|}{2} \in [0,1]$$

Balance in screen design was achieved by providing an equal weight of screen elements, left and right, top, and bottom. Balance was computed as the difference between total weighting of components on each side of the horizontal and vertical axis.

2. Equilibrium

$$Equilibrium = 1 - \frac{|Equilibrium_X| + |Equilibrium_Y|}{2} \in [0,1]$$

Equilibrium on a screen was accomplished through centering the layout itself. Equilibrium was computed as the difference between the center of mass of the displayed elements and the physical center of the screen.

3. Symmetry

$$Symmetry = 1 - \frac{|SymmetryVer| + |SymmetryHor| + |SymmetryRad|}{3} \in [0,1]$$

Symmetry was axial duplication where a unit on one side of the centerline was exactly replicated on the other side. There were three types of symmetry, which were vertical symmetry, horizontal symmetry, and radial symmetry. Symmetry, by definition, was the extent to which the screen is symmetrical in three directions: vertical, horizontal, and diagonal.

4. Sequence

$$Sequence = 1 - \frac{\sum_{j=UL,UR,LL,LR} |q_j - v_j|}{8} \in [0,1]$$

Sequence in design referred to the arrangement of objects in a layout in a way that facilitated the movement of the eye through the information displayed. Sequence, by definition, was a measure of how information in a display was ordered in relation to a reading pattern that was common.

5. Rhythm

$$Rhythm = 1 - \frac{|Rhythm_X| + |Rhythm_Y| + |Rhythm_{Area}|}{3} \in [0,1]$$

Rhythm was accomplished through variation of arrangement, dimension, number, and form of the elements. The extent to which rhythm was introduced into a group of elements depends on the complexity. Rhythm, by definition, was the extent to which the objects are systematically ordered.

6. Order and Complexity

$$Order_Complexity = \frac{\sum_{i}^{5} M_i}{5} \in [0,1]$$

The measure of order was written as the sum of the above measures for a layout. The opposite pole on the continuum was complexity. The scale created might also be considered as a scale of complexity, with extreme complexity at one end and minimal complexity (order) at the other.

Table 4 showed the object models of the twelve Mandarin learning web page interfaces. The object models of the Mandarin learning web page interfaces were manipulated through image processing by using Photoshop program. These object models were used as modeling of the web page interfaces and showed the objects on each of the Mandarin learning web page interface clearly. While Table 5 showed the results of aesthetic values of twelve web pages used in our research. These values were between 0 (worst) and 1 (best). Twelve Mandarin learning web pages were developed and tested according to the manipulated aesthetic measurement. They were four main pages, four learning pages and four exercise pages. These were arranged into four groups that will be tested later on. Group 1 was GUIs with highest aesthetic values, while Group 4 was GUIs with the lowest aesthetic values.

Table 4: Object Model of Twelve Mandarin Learning Web Page Interfaces

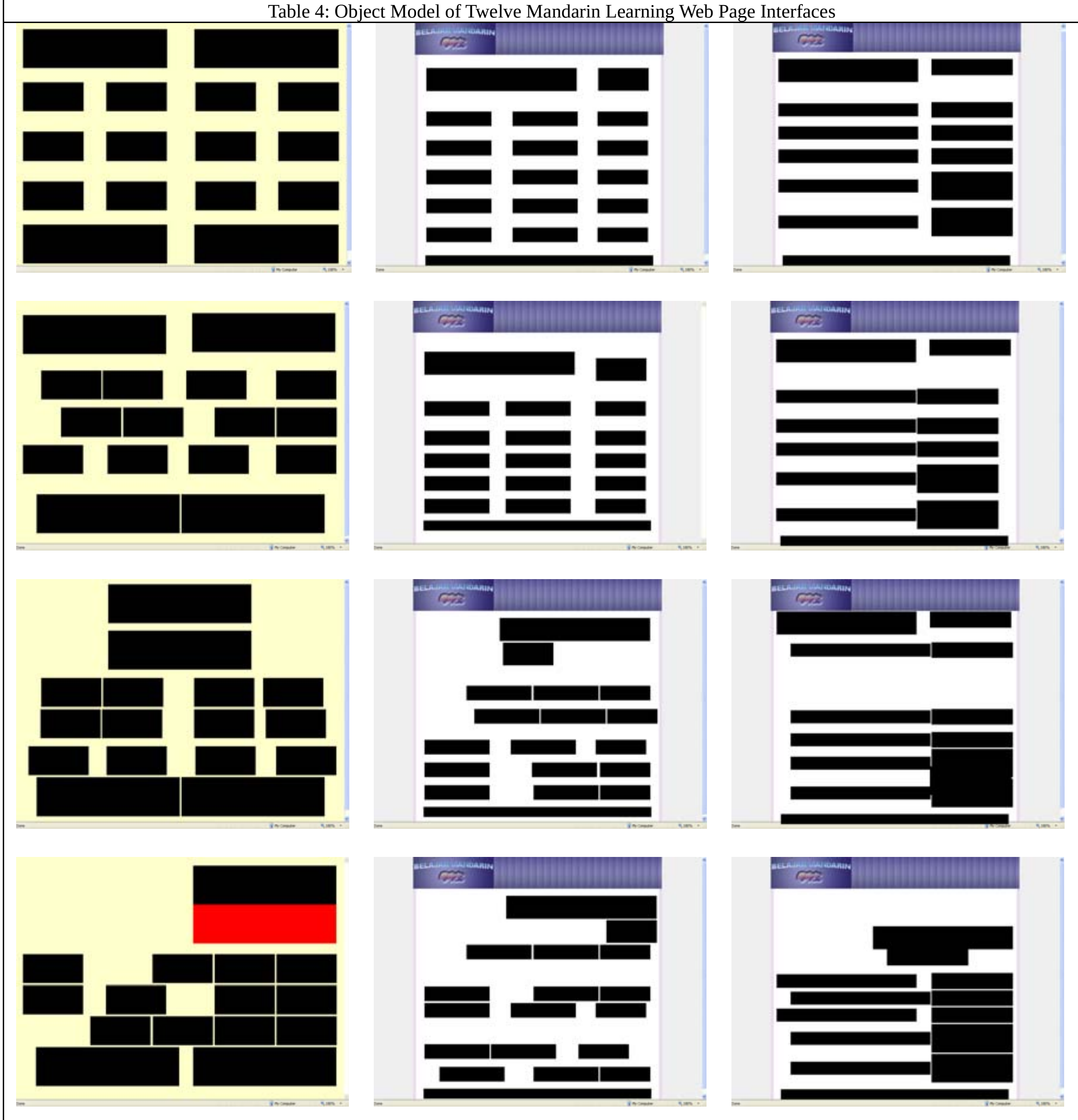

Table 5: Results of Aesthetic Values (avs) of Mandarin Learning Web Pages by Using AMA

| | | | | | |
|---|---|---|---|---|---|
| **Main Page (Group 1)** | 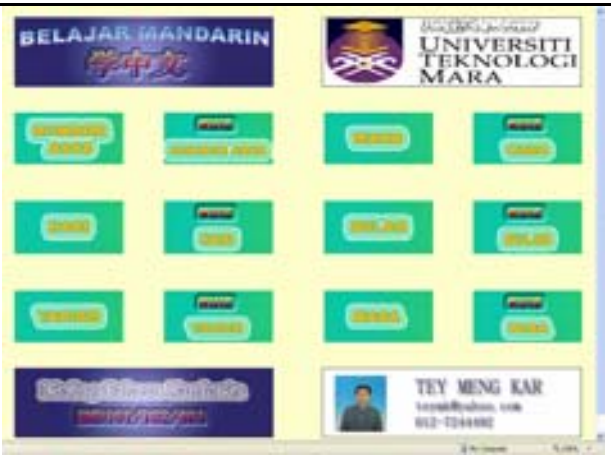<br>Balance 0.9445<br>Equilibrium 0.9991<br>Symmetry 0.9013<br>Sequence 1.0000<br>Rhythm 0.9085<br>**Aesthetic value (av) 0.9507** | **Learning Page (Group 1)** | 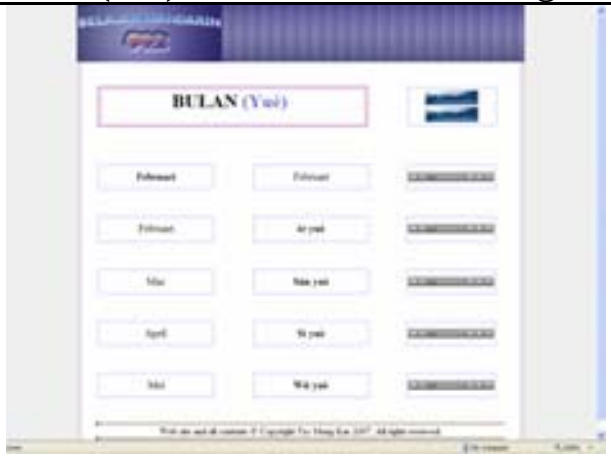<br>Balance 0.6558<br>Equilibrium 0.9954<br>Symmetry 0.6062<br>Sequence 0.7500<br>Rhythm 0.6663<br>**Aesthetic value (av) 0.7347** | **Exercise Page (Group 1)** | 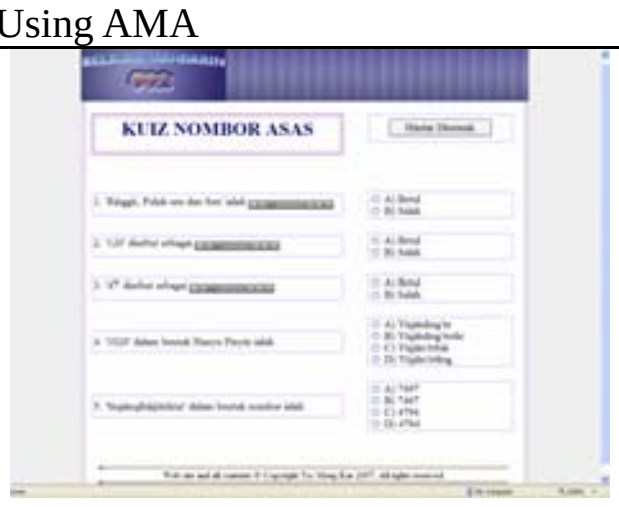<br>Balance 0.8054<br>Equilibrium 0.9965<br>Symmetry 0.4402<br>Sequence 0.7500<br>Rhythm 0.5592<br>**Aesthetic value (av) 0.7103** |
| **Main Page (Group 2)** | 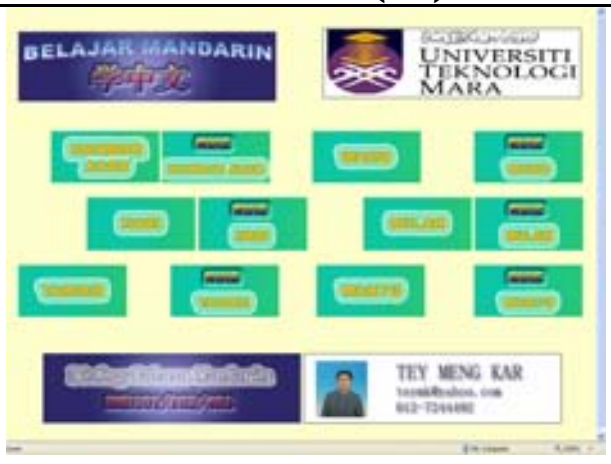<br>Balance 0.9369<br>Equilibrium 0.9990<br>Symmetry 0.8234<br>Sequence 1.0000<br>Rhythm 0.8700<br>**Aesthetic value (av) 0.9259** | **Learning Page (Group 2)** | 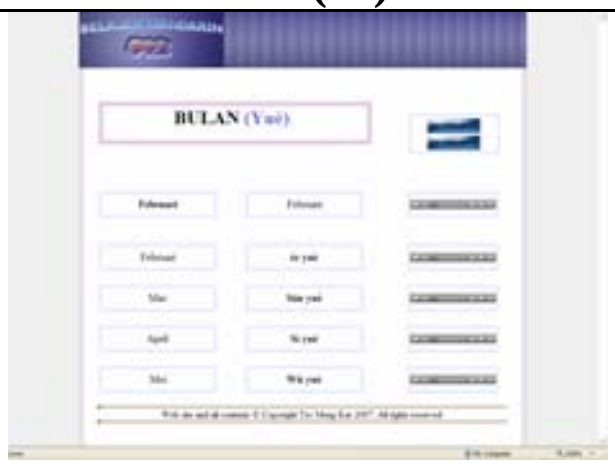<br>Balance 0.5784<br>Equilibrium 0.9945<br>Symmetry 0.4161<br>Sequence 0.7500<br>Rhythm 0.4917<br>**Aesthetic value (av) 0.6461** | **Exercise Page (Group 2)** | 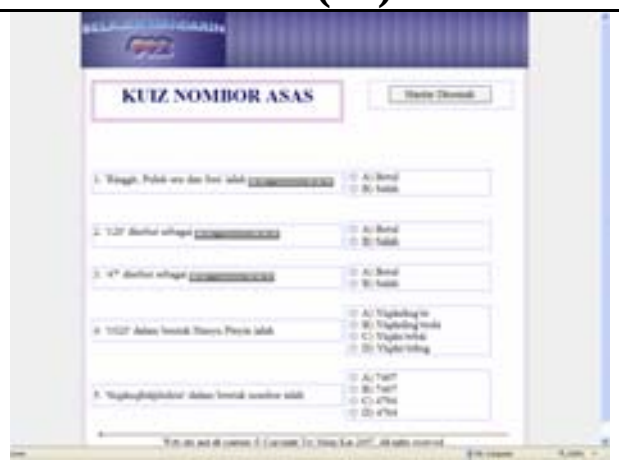<br>Balance 0.6911<br>Equilibrium 0.9932<br>Symmetry 0.3796<br>Sequence 0.7500<br>Rhythm 0.4331<br>**Aesthetic value (av) 0.6494** |
| **Main Page (Group 3)** | 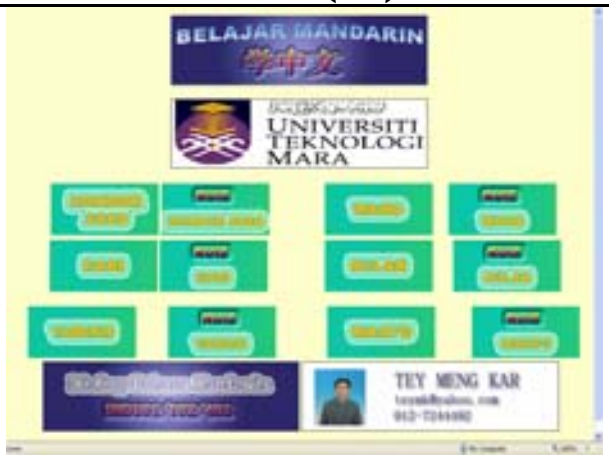<br>Balance 0.7656<br>Equilibrium 0.9960<br>Symmetry 0.4958<br>Sequence 0.6250<br>Rhythm 0.5324<br>**Aesthetic value (av) 0.6830** | **Learning Page (Group 3)** | 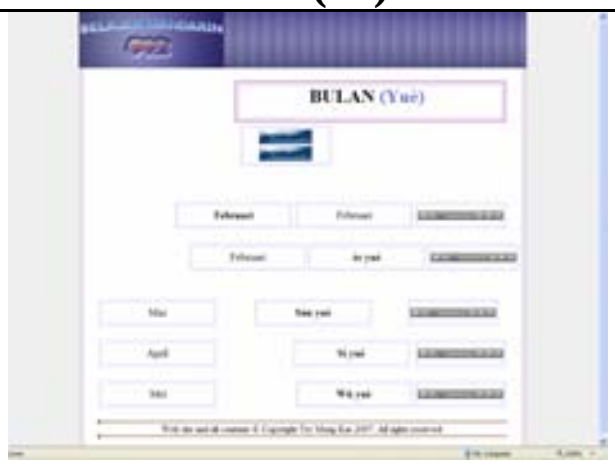<br>Balance 0.5309<br>Equilibrium 0.9935<br>Symmetry 0.4555<br>Sequence 0.5000<br>Rhythm 0.4870<br>**Aesthetic value (av) 0.5934** | **Exercise Page (Group 3)** | 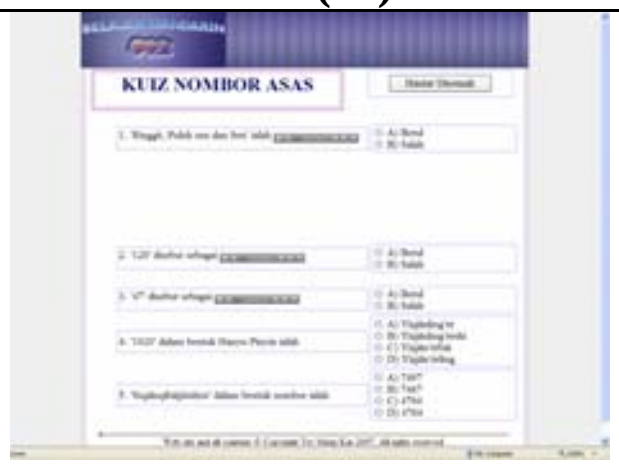<br>Balance 0.5944<br>Equilibrium 0.9913<br>Symmetry 0.4515<br>Sequence 0.5000<br>Rhythm 0.3459<br>**Aesthetic value (av) 0.5766** |
| **Main Page (Group 4)** | 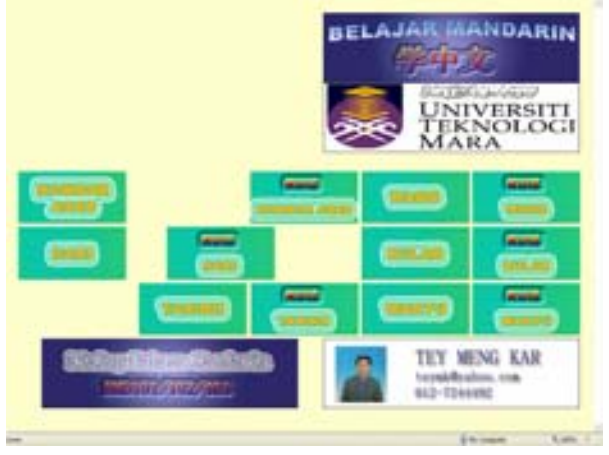<br>Balance 0.5674<br>Equilibrium 0.9918<br>Symmetry 0.2689<br>Sequence 0.3750<br>Rhythm 0.2258<br>**Aesthetic value (av) 0.4858** | **Learning Page (Group 4)** | 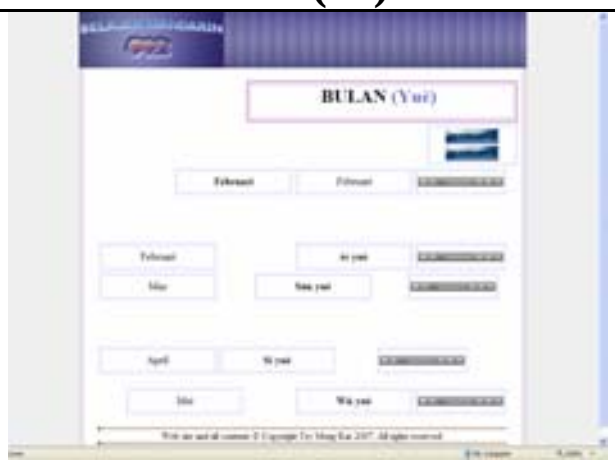<br>Balance 0.5411<br>Equilibrium 0.9934<br>Symmetry 0.3399<br>Sequence 0.3750<br>Rhythm 0.3511<br>**Aesthetic value (av) 0.5201** | **Exercise Page (Group 4)** | 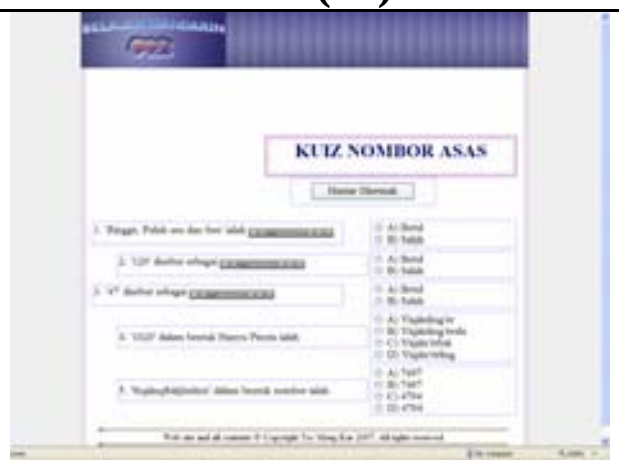<br>Balance 0.3296<br>Equilibrium 0.9859<br>Symmetry 0.3421<br>Sequence 0.5000<br>Rhythm 0.3134<br>**Aesthetic value (av) 0.4942** |

### 3.2. User's Perception of Aesthetics on Web Page Interfaces and Aesthetics Relevance with Mandarin Learning

With the objectives in justifying that there is a significant correlation of the aesthetic aspect with its relevance to Mandarin learning, the Mandarin learning web pages created were used. In this research, non-random sampling method was used to gather the information needed.

Students from two classes, namely Class A and Class B, who were taking Mandarin level I course (BMD 401) at University of Technology MARA Terengganu branch were involved. Class A was from Faculty of Accountancy and Class B was from Faculty of Business Management and Faculty of Office Management. Both classes were taught by the same Mandarin instructor.

The research was conducted at the beginning of the semester. The students from these two classes were assigned into four groups as showed in Table 6. Lessons were conducted in a computer lab using the Mandarin learning web pages that were uploaded in the Mandarin instructor's e-portfolio (http://www1.tganu.uitm.edu.my/teymk/teaching/number/mandarin.html) and followed the directions given.

Table 6: Respondents' Background

| **Bachelor in** | **Group** | | | | |
|---|---|---|---|---|---|
| | **1** | **2** | **3** | **4** | **Total** |
| Accountancy | 5 | 5 | 5 | 4 | 19 |
| Office Management & Technology | 10 | 9 | 5 | 10 | 34 |
| Business Management | 0 | 1 | 5 | 1 | 7 |
| **Total** | **15** | **15** | **15** | **15** | **60** |

Every student was allocated a computer to use in the computer lab. The topic chosen was "Number in Mandarin". Thus, all the contents of these learning pages used were relevant to this topic. The contents of the web page were relevant to number as learning materials. Concerning the instructional events, first, a learning page on the topic "General Number" was shown. This was then followed by learning pages around this topic, which were Days, Months and Dates in Mandarin. Then, the students were required to try the relevant exercise pages. Every group (group 1, group 2, group 3, and group 4) accessed different Mandarin learning web pages with the same contents.

The questionnaire used in this research was self-developed. It was mainly to gain students' perceptions on the ranking of web pages used as well as students' perceptions on the relevance of aesthetics to their Mandarin learning experiences. The students were asked to judge the aesthetic aspects of the pages shown by using Likert scale from 1-5 where 1 means the worst and 5 means the best. The students were asked to rate forty items altogether on five important parts in the questionnaire given. There were eight items in aesthetic part on Main page, Learning page and Exercise page individually. Then, eight items in aesthetic and its relevance to Mandarin learning part on number as well as eight items in aesthetic and its relevance to Mandarin learning part on general.

## 4. Findings and Discussions

Table 7 showed the comparisons between the ranking of results using self-developed Aesthetics-Measurement Application (AMA) and questionnaires distributed to the students in gaining their views on the aesthetics of web pages as well as the relevance of aesthetics with Mandarin learning. The results of the application showed that web pages of group 1 were the most aesthetic and followed by group 2, group 3 and the group 4 was the most unaesthetic ones. The findings from the students' perceptions on main pages, learning pages, and exercise pages showed that web pages of group 1 were viewed as the most aesthetic, and followed by group 3, group 2 and group 4 was the most unaesthetic ones. There was a reverse in the positions of ranking of the aesthetics measured by using the application and students' perceptions.

Students were asked on the aesthetics with their Mandarin Numbers' learning experience in particular. Means of the group 1 was the highest, followed by group 3; group 2 and mean of the group 4 was the lowest. Again, we noticed that there was an overturn in the positions of ranking on the aesthetic values in relations to students' perceptions.

While on the aspect of aesthetics in relevance to Mandarin in general, the findings showed that, means of the group 1 was the highest, followed by group 3, group 4 and mean of the group 2 was the lowest. The results did not follow the ranking as anticipated.

The ranking of group 1 stayed at the first position in all aspects; therefore, it supported the view that the more aesthetic a web page, the more it motivates students to learn Mandarin. At the same time, the ranking of group 4 remained the last position in all aspects, thus, it also supported our assumption that aesthetics is relevant to Mandarin learning.

As a result, it is appropriate to conclude that the findings confirmed our scrutiny that aesthetics affects students' learning. Even though there were some difference with the positions of Groups 3 and 4 to our anticipation, that might be due to some shortcomings of the designs, yet, as a whole, it was still quite convincing to claim that aesthetics played its role in affecting students' learning motivation.

However, the findings on students' perception on aesthetics and learning in general were not in line to our expectation. There might be some other factors that were undermined and needed to be investigated that might have affected students' views.

Relating to the discussions of findings above, Table 8 showed the result of ANOVA test (F–Test) used. For aesthetics of main pages, with F value = 4.381, significant value = .008 (smaller than 0.05), it was significant. Thus, it confirmed the claim that the findings of students' perceptions on the aesthetics were in congruent to the measurements of our self-developed Aesthetics-Measurement Application (AMA). The application can be used to measure the aesthetics of web pages. For aesthetics of learning pages, with F value = 3.718, significant value = .016 (smaller than 0.05), it was significant. Thus, it did support our claim that our self-developed Aesthetics-Measurement Application (AMA) measured as the students' views. While aesthetics of exercise pages, with F value = 7.272, significant value = .000 (smaller than 0.05), it was significant. Thus, it confirmed the claim that the findings of students' perceptions on the aesthetics were in congruent to the measurements of our self-developed Aesthetics-Measurement Application (AMA). The application can be used to measure the aesthetics of web pages.

Relevance of aesthetics and learning Mandarin numbers, with F value = 4.615, significant value = .006 (smaller than 0.05), it was significant. Thus, it confirmed the claim that aesthetics did affect students' Mandarin learning. For aesthetics and learning in general, with F value = 2.175, significant value = .101 (greater than 0.05), it was not significant. Thus, this research did not confirm the claim that aesthetics affected students' learning in general.

To examine the significance in the comparison among means of four groups, ANOVA test (T–Test) was used. Table 9 showed the result of comparison. With F value = 5.197, significant value = .003 (smaller than 0.05), it was significant. Therefore, the respondents from different groups held different aesthetic view and learning experiences. In some cases, it did confirm that more aesthetic of web pages (main page, learning page, and exercise page), the more students were motivated to learn Mandarin. As the result, the findings did support our assertion that learning by using GUIs is affected by the aesthetics of the GUIs.

Table 7: Comparison between Ranking of the Results of AMA and Users' Perception

| **GROU** | **Application (ranking)** | | | **Questionnaire (ranking), (Aesthetic)** | | | **Questionnaire (ranking), (Aesthetic & learning)** | |
|---|---|---|---|---|---|---|---|---|
| | **Web page** | | | **Web page** | | | | |
| | **Main** | **Learning** | **Exercise** | **Main** | **Learning** | **Exercise** | **Number** | **General** |
| **1** | 0.9507 (1) | 0.7347 (1) | 0.7103 (1) | 4.5417 **(1)** | 4.5833 **(1)** | 4.5583 **(1)** | 4.3667 **(1)** | 4.3733 **(1)** |
| **2** | 0.9259 (2) | 0.6461 (2) | 0.6494 (2) | 3.8333 **(3)** | 4.1167 **(3)** | 3.9667 **(3)** | 3.7267 **(3)** | 3.8933 **(4)** |
| **3** | 0.6830 (3) | 0.5934 (3) | 0.5766 (3) | 4.1917 **(2)** | 4.1333 **(2)** | 4.5250 **(2)** | 4.1867 **(2)** | 4.3000 **(2)** |
| **4** | 0.4858 (4) | 0.5201 (4) | 0.4942 (4) | 3.6833 **(4)** | 3.8000 **(4)** | 3.6750 **(4)** | 3.7000 **(4)** | 3.9800 **(3)** |

Table 8: ANOVA Tests (F-Test) for Five Hypotheses in This Study

| | | **df** | **F** | **Sig.** |
|---|---|---|---|---|
| **Aesthetic of main page** | Between Groups<br>Within Groups | 3<br>56 | 4.381 | **.008** |
| **Aesthetic of learning page** | Between Groups<br>Within Groups | 3<br>56 | 3.718 | **.016** |
| **Aesthetic of exercise page** | Between Groups<br>Within Groups | 3<br>56 | 7.272 | **.000** |
| **Aesthetic and learning (number)** | Between Groups<br>Within Groups | 3<br>56 | 4.615 | **.006** |
| **Aesthetic and learning (general)** | Between Groups<br>Within Groups | 3<br>56 | 2.175 | **.101** |

Table 9: ANOVA Test (T-Test) - Comparing Means between 4 Groups

| | **df** | **F** | **Sig.** |
|---|---|---|---|
| Between Groups<br>Within Groups<br>Total | 3<br>56<br>59 | 5.197 | **.003** |

## 5. Implications, Suggestions and Conclusion

These findings showed an optimistic view in the aesthetic aspect with its relevance to Mandarin learning. The result of the findings showed that the ranking for group 1 and 4 remained the same as the measurements using the self-developed Aesthetics-Measurement Application (AMA). From the similar research done in past, it was also found that users' perceptions were congruent with the aesthetics values measured using application [14]. Thus, the calculations and technique of aesthetic measurement for screen layout in this research can be used.

However, it was that group 2 ranked the third place from the students' responses. It was incongruent to our expectation. This happened to group 3 that ranked the second place. It shows that there might be an issue of subjectivity here. It may be caused by the respondents' inconsistency in answering the questionnaire. Other factor might be due to the limited options given to the students. The students were required to pick their choices from Likert scale in 1-5. The differences among five might be not enough. It is suggested that a scale of 10 should be used in order to see a clearer picture.

Respondents agreed that learning numbers in Mandarin with aesthetic web pages could increase their motivation in learning. It is congruent with our assertion that the more aesthetic the web pages, the more it could motivate students to learn Mandarin. However, this research finding could not support our claims that aesthetics have an effect on both Mandarin learning and learning in general. There might be some other reasons that affected students' views, such as the interactivity of an exercise page, time limit in viewing, etc. Thus, further research could be done where all these factors will be taken into consideration in the research design.

As a conclusion, the self-developed Aesthetics-Measurement Application (AMA) can be considered as a viable tool in evaluating the aesthetics of an interface. Six aesthetic measures were used only in this research, which consisted of balance, equilibrium, symmetry, sequence, rhythm, as well as order and complexity in order to demonstrate that aesthetics can be measured objectively for multi-screen interfaces. Consequently, other influential factors were not taken into our measurement. Thus, this opens further research possibility of taking in other aesthetics measurers to develop a more powerful aesthetics-measuring tool. Furthermore, measuring aesthetic of interface can also include shapes of objects, colors of objects, texts, frames, and background of screen. This research can also be extended to the measuring of aesthetics of multi windows, multi document, and multi pane interfaces.

Consequently, measurement of aesthetics can be viewed as an important task that should not be disregarded. This is because aesthetics of interface would influence usability, acceptability, learn-ability, comprehensibility, and productivity. It is thus suggested that the more aesthetic a web page is, the more useful the web page can be. In advocating the contentions of these researchers that the more aesthetic the web pages, the more it could motivate students to learn Mandarin particularly and for learning in general [14]. Nevertheless, more research should be done to validate the findings. Replication of this research at different settings and branch campuses will be appropriate. On the other hand, aesthetic quality and attractiveness aspect in the e-portfolio was an aspect that ought not to neglect; aesthetic aspect needs to be attended by the instructors and it is important that instructors have to understand what is pleasing to the students [2]. Instead of preparing long lists of learning materials in the web pages of the E-portfolio, which are not appealing to the students, certainly it would not serve its purpose. Therefore, it is important to develop more aesthetically pleasing web pages of the E-portfolio to make learning process successfully.

In conclusion, the findings are valuable to all the foreign language instructors generally in their endeavors in using technology to enhance their foreign language classroom as well. Web pages should be designed in such a manner that the aesthetic values are taken into concern. By then, all kinds of instructional technology used in foreign language classrooms have to maintain a necessary aesthetics standard in bringing supportive environment for Mandarin learner particularly and for all foreign language students generally. In addition, further research also can be done to investigate the relationship between aesthetics and learning for purpose to develop more pleasing layout of web pages of the E-portfolio as the learning material to supplement Mandarin Learning as well as for other language learning.

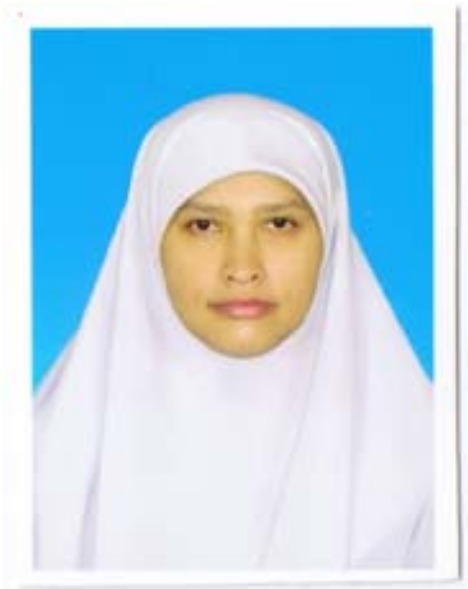

**Jasni Mohamad Zain** received her Bachelor degree in Computer Science from University of Liverpool, England, UK in 1989; PGCE Mathematics from Sheffield Hallam University, England, UK in 1994; M.E. degree from Hull University, England, UK in 1998 and PhD from Brunel University, West London, UK in 2005. She currently holds the post as the Director of the Center of Information Technology and Communication, University Malaysia Pahang. She is currently a lecturer in Faculty of Computer Science and Software Engineering, University Malaysia Pahang. She has been actively presenting papers in national and international conferences. Her research interests include Image Processing as well as Data and Network Security.

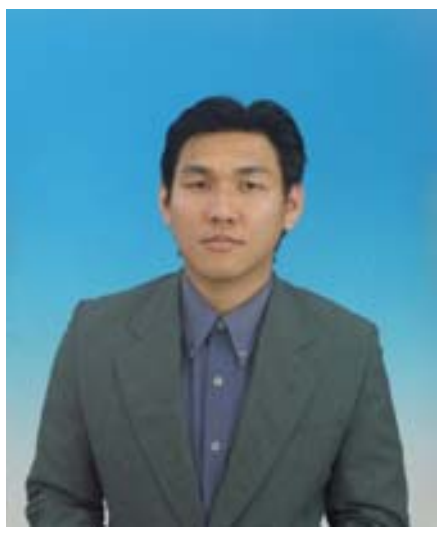

**Mengkar Tey** currently is pursuing his master degree in Software Engineering in University Malaysia Pahang under supervision Dr. Jasni Mohamad Zain and Mr. Yingsoon Goh. He has experiences in teaching Mandarin as the third language to non-native learners in MARA University of Technology, Malaysia for almost 2 years. His research interests are on the web page development, image processing, as well as Chinese character handwriting recognition.

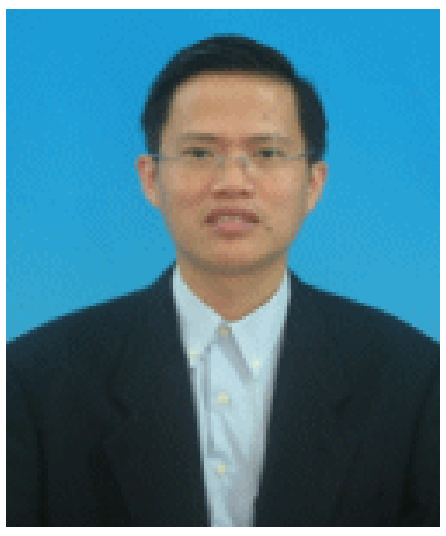

**Yingsoon Goh** currently teaches Mandarin as the third language to non-native learners in MARA University of Technology, Malaysia. He has experiences in teaching Mandarin at primary, secondary, and tertiary level for almost 17 years. He has been actively presenting papers in national and international conferences. His research interests are on the use of educational technology in Mandarin teaching and learning, as well as web-based instruction.